%% file: main.tex
\documentclass[letterpaper, 10 pt, conference]{ieeeconf}
\IEEEoverridecommandlockouts                            
\input{support/preamble}

\input{support/macros}
\title{\LARGE \bf Parameter Robustness in Data-Driven Estimation of Dynamical Systems}

\author{Ayush Pandey \thanks{This work was supported by the CITRIS-UC Institute Grant \#2023-025.}% <-this % stops a space
\thanks{A. Pandey is with the Electrical Engineering department at the University of California, Merced, CA 95343 USA. {\tt\small Email: ayushpandey@ucmerced.edu}
Paper Code: {\tt \small https://github.com/ayush9pandey/robest}
}
}

\begin{document}

\maketitle
\thispagestyle{empty}
\pagestyle{empty}

%%%%%%%%%%%%%%%%%%%%%%%%%%%%%%%%%%%%%%%%%%%%%%%%%%%%%%%%%%%%%%%%%%%%%%%%%%%%%%%%
\begin{abstract}
We study the robustness of system estimation to parametric perturbations in system dynamics and initial conditions. We define the problem of sensitivity-based parametric uncertainty quantification in dynamical system estimation. The main contribution of this paper is the development of a novel robustness metric for estimation of parametrized linear dynamical systems with and without control actions. For the computation of this metric, we delineate the uncertainty contributions arising from control actions, system dynamics, and initial conditions. Furthermore, to validate our theoretical findings, we establish connections between these new results and the existing literature on the robustness of model reduction. This work provides guidance for selecting estimation methods based on tolerable levels of parametric uncertainty and paves the way for new cost functions in data-driven estimation that reward sensitivity to a desired subset of parameters while penalizing others.
\end{abstract}

%%%%%%%%%%%%%%%%%%%%%%%%%%%%%%%%%%%%%%%%%%%%%%%%%%%%%%%%%%%%%%%%%%%%%%%%%%%%%%%%
\section{Introduction}
\input{sections/introduction}
\section{Problem Formulation}
\input{sections/problem-setup}
\section{Preliminary Results}
\input{sections/preliminaries}
\section{Results}
\input{sections/thm-A}
\input{sections/results}
\section{Application Example: A 2D LPV System}
\input{sections/applications}
\section{Conclusions and future work}
\input{sections/conclusions}
% \section*{Acknowledgments}
% The author thanks Alex Frias, Ebonye Smith, Gireeja Ranade, and S. Shailja for many helpful conversations related to the topics discussed in this paper. 
% 
% Generative AI statement: ChatGPT~\cite{achiam2023gpt} was used to improve the syntax and grammar of several paragraphs in the manuscript. GitHub Copilot~\cite{copilot} was used to generate boilerplate code with its autocomplete feature, which was verified by the author. 

%%%%%%%%%%%%%%%%%%%%%%%%%%%%%%%%%%%%%%%%%%%%%%%%%%%%%%%%%%%%%%%%%%%%%%%%%%%%%%%%

\bibliography{bibtex/bib/IEEEabrv.bib,bibtex/bib/references.bib}{}
\bibliographystyle{IEEEtran}
% \section*{Appendix: Proofs of Integrals}
% \input{sections/appendix-integrals}
\input{sections/appendix_a2b2}
\input{sections/appendix_baseline}
\end{document}

%% file: support/preamble.tex
\usepackage{cite}  
\usepackage{graphics} % for pdf, bitmapped graphics files
\usepackage{amsmath} % assumes amsmath package installed
\usepackage{amssymb,mathtools}  % assumes amsmath package installed
\usepackage{xcolor}
\usepackage{pgfplots,subcaption}
\usepackage[hidelinks]{hyperref}
\usepackage{verbatim}
\pgfplotsset{width=4.5cm,compat=1.9}
\usepgfplotslibrary{fillbetween}

\newsavebox{\measurebox}

\newcommand{\norm}[1]{\left\lVert#1\right\rVert}
\def\abs#1{\left\lvert#1\right\rvert}

\usepackage{amsthm}
\usepackage{accents}
\usepackage{relsize}

\newtheorem{theorem}{Theorem}

\newtheorem{lemma}{Lemma}

\theoremstyle{definition}

\theoremstyle{remark}

\theoremstyle{remark}

\newenvironment{nalign}{
    \begin{equation}
    \begin{aligned}
}{
    \end{aligned}
    \end{equation}
    \ignorespacesafterend
}

%% file: sections/introduction.tex
Dynamical systems modeling and control methods have evolved with the rise of data-driven paradigms, enabled by rapid, large-scale access to data from complex physical systems. Estimation and control, the two fundamental problems in control theory, have significantly benefited from easily applicable data-driven methods. As these methods continue to advance, system designers must select the most appropriate technique to estimate system dynamics from an expanding range of options. We study this problem in the context of parametric robustness of system estimation. We start by discussing the existing literature on robustness quantification in estimation and control.
\subsection{Data-driven estimation and control}
Various data-driven system estimation and control methods have been developed in the control literature. Among these, the Dynamic Mode Decomposition (DMD)~\cite{schmid2010dynamic} has emerged as a popular choice to decompose complex system dynamics into identifiable modes that describe the system dynamics. DMD provides a linear estimate of discrete-time system dynamics from time-series snapshots and has been extended for control as DMD with control (DMDc) to handle actuated inputs~\cite{proctor2016dynamic}, for systems with sensor noise~\cite{dawson2016characterizing}, and other similar settings. Recently, a Python tool called pyDMD~\cite{demo2018pydmd, ichinaga2024pydmd} was developed that encompasses most of these DMD related tools in a user-friendly package. A related development in data-driven system estimation is the Sparse Identification of Nonlinear Dynamics (SINDy) method for continuous-time system dynamics. SINDy uses sparse regression to discover governing equations from time-series data, yielding interpretable models~\cite{brunton2016discovering} 
% which have been successfully applied for the estimation of fluid dynamic models~\cite{strom2022near, loiseau2018constrained}, flight control~\cite{kaiser2018sparse}, among other physical system examples. 
These methods are founded in an operator theoretic framework that applies Koopman theory~\cite{brunton2021modern} for finding embeddings of nonlinear dynamics using a high-dimensional linear operator.
Beyond these operator theoretic data-driven methods, various model-free and learning-based control methods have become increasingly common in control applications. For example, reinforcement learning algorithms are now the go-to methods to learn close-to-optimal control policies from data for applications in robotics and beyond. RL algorithms can be iteratively tuned to achieve significant gains in performance (lower tracking error, better disturbance rejection), greater flexibility (adapting to new operating regimes without re-deriving equations), and improved computational efficiency (enabling real-time implementation) but often lack formal guarantees. Towards that end, researchers are exploring the integration of neural network function approximators with physics-based models to improve generalization and interpretability~\cite{darabi2025combining}. Although this is still an active and open area of research, a recent example~\cite{champion2019data} includes the use of deep autoencoders to discover coordinates for SINDy to inform the control decisions using a latent space where the dynamics are sparsely described by SINDy. Hybrid methods leverage neural networks to handle complex feature extraction (difficult in manual basis selection) while constraining the model structure using first-principles physics-based modeling. Along this line, the area of physics-informed neural networks, or gray-box modeling is growing rapidly and is impacting the ways in which control theorists implement system estimation using data~\cite{darabi2025combining,lee2024robot}.
Despite the progress, the problem of guarantees on robustness and safety~\cite{martin2023guarantees,pacti-paper,ames-guarantees,chuchu-fan,nik-matni} of estimation and control is actively studied. 
% There is significant interest from the control theorists in bridging data-driven methods with formal methods and physics-informed approaches to embed safety and robustness guarantees in data-driven pipelines.
\subsection{Robustness quantification}
Both empirical and analytical benchmarking of robustness are common. Empirical approaches include multiple runs for the same system or dataset that can be compared and used for benchmarking, and visual illustrations using time-series plots or phase-plane trajectories. On the other hand, analytical approaches include the computation of signal to noise (or disturbance) ratio, or the computation of metrics such as the mean squared error (MSE). Classical control theory offers frequency-domain robustness metrics such as the $\mathbb{H}_\infty$ norm (the worst-case gain across frequencies) and $\mu$-analysis~\cite{doyle2013feedback}. However, these frequency-domain metrics are less applicable in data-driven settings. For example, a recent study~\cite{gupta2023direct} demonstrates the use of fixed-structure controllers designed directly from frequency response data, minimizing $\mathbb{H}_2$ or $\mathbb{H}_{\infty}$ costs without explicit state-space models~\cite{Schuchert2023}. Despite such efforts, data-driven research predominantly emphasizes time-domain metrics due to inherent limitations like unclear error bounds, nonlinearities, high-dimensional models, and conservative robustness constraints. 
\subsection{The need for sensitivity-driven robustness}
In this paper, we take an approach motivated by sensitivity analysis, an area that has historically been integral to control theory but remains underexplored due to the success of frequency-domain methods. The key idea is to analyze the partial derivative of a dynamical quantity of interest with change in the model parameters. This provides a direct measure of how specific parameters influence the behavior of the system over time. This facilitates seamless integration into data-driven pipelines. More importantly, with this approach, we can explore robustness measures against specific parameters of interest for a precise analysis that is often of interest in the design of physical systems. For example, in a rotor-driven robotic system, one may be particularly interested in ensuring stable hovering performance despite variations in motor constants or aerodynamic coefficients. Another motivating example comes from engineered biological systems, where a protein's concentration might be designed to respond strongly to certain dynamically varying parameters, while remaining robustly insensitive to fluctuations in other parameters within a specified range. Such precise, parameter-specific robustness analyses are challenging but critical, and our approach seeks to enable these capabilities within modern data-driven modeling and control pipelines.

\textbf{Contributions:} In this paper, we express the problem of robustness of error in data-driven system estimation and derive an easily computable bound for this robustness metric for linear systems with control. Specifically,
\begin{enumerate}
    \item We define a new robustness metric for data-driven estimation of parametrized linear dynamical systems with and without actuation. 
    \item For the computation of our new robustness metric, the main theorem in this paper gives a bound to efficiently quantify the robustness in the estimation of parameterized system dynamics, without solving for system trajectories.
    \item To prove the validity of our results, we draw the equivalence of the main theorem in special cases to already known results in the literature. 
    % \item The applications of our robustness metric provide a way forward in choosing estimation methods according to the levels of parametric uncertainties that may be tolerated in the estimated system.
\end{enumerate}
\textbf{Paper outline:} We start by formally defining the problem in the next section. Then, in Section~\ref{prelims}, we present important preliminaries to derive the main result of the paper given in Section~\ref{results}. 
We discuss a linear parameter varying (LPV) system example in Section~\ref{applications}. 

%% file: sections/problem-setup.tex
\label{setup}
\subsection{System description}
We consider the problem of estimation of a parametrized linear dynamical system with control inputs, 
\begin{nalign}
\dot{x} &= A(\theta)x + B(\theta)u,\\
y &= C(\theta)x + D(\theta)u \label{eq:sys}
\end{nalign}
where $x \in \mathbb{R}^n$ is the state vector, $A \in \mathbb{R}^{n\times n}$ is a real-valued system matrix, $B \in \mathbb{R}^{n\times k}$ is the control matrix for the system with $k$ inputs, $u \in \mathbb{R}^k$, $y \in \mathbb{R}^{m}$ is the output vector consisting of $m$ measurements with the output matrix $C \in \mathbb{R}^{m \times n}$, and $D \in \mathbb{R}^{m \times k}$ is the feedforward matrix for the system. In the most general case of parametric dependence of the system dynamics, all matrices can be functions of a parameter vector $\theta \in \mathbb{R}^{p}$, which consists of all $p$ parameters of the system. An estimation of this system is written as
\begin{nalign}
\dot{\tilde{x}} &= \tilde{A}(\theta)\tilde{x} + \tilde{B}(\theta)u,\\
\tilde{y} &= \tilde{C}(\theta)\tilde{x} + \tilde{D}(\theta)u \label{eq:est_sys}
\end{nalign}
where the $\tilde{.}$ represents the corresponding estimated state variables while the control inputs $u$ remain the same since these are usually external forces, which will be applied in the same way to the estimated system as well. Note that we assume here that the estimated dynamical system is also parametrized. This may not be the case for many `off-the-shelf' data-driven methods for system estimation. However, enforcing such a structure on estimated dynamics is possible and has been used in the literature~\cite{baddoo2023physics,lee2022structure} for many data-driven system estimation methods for parametrized systems. Thus, the assumption of this structure is not severely limiting. 
\subsection{Robustness metric}
The key quantity that we are interested in is the quantification of robustness to parametric uncertainties. For the system estimation problem, we aim to compute a metric for the change in estimation error as model parameters (or initial conditions) vary. This gives a robustness estimate for the estimation method. By adapting a similar robustness metric from prior work~\cite{pandey2023robustness}, we define a robustness distance that measures how the estimation error changes as a model parameter $\theta_i^*$ varies as
\begin{equation}
    d_R = \mathlarger{\mathlarger{\sum}}_{i = 1}^{p}\frac{\theta_i^*}{\norm{\text{err}(t, \theta_i^*)}}\cdot \norm{\left.\frac{\partial \text{err}}{\partial \theta_i}\right\rvert_{\theta_i = \theta_i^*}},
    \label{robustness_distance}
\end{equation}
where $\text{err}(t,\theta_i^*)$ is the non-zero error in estimation ($y - \tilde{y}$) for $t > 0$. Using the distance $d_R$, we define the robustness metric of estimation error that lies between $[0,1]$ as
\begin{equation}
    \label{robustness_metric}
    R = \frac{1}{1 + d_R} = \left(1 + \mathlarger{\mathlarger{\sum}}_{i = 1}^{p}\frac{\theta_i^*\norm{\left.\frac{\partial \text{err}}{\partial \theta_i}\right\rvert_{\theta_i = \theta_i^*}}}{\norm{\text{err}(t, \theta_i^*)}}\right)^{-1}.
\end{equation}
When the distance $d_R = 0$ (estimation error did not change as parameters varied), the robustness metric is at its maximum $R = 1$, and vice versa if the estimation error is highly sensitive $R$ will be closer to 0 (less robust).
\subsection{Augmented system dynamics}
% The main idea in the problem formulation is the description of an augmented system dynamics. 
By augmenting the ground truth system matrices with the estimated matrices, we obtain a form that enables the computation of the robustness metric. For this, we define the augmented state variables and matrices by using the $\bar.$ notation as follows:
\begin{align}
\bar{x} &= \begin{bmatrix} x \\ \tilde{x} \end{bmatrix}, \quad
\bar{A} = \begin{bmatrix} A & 0 \\ 0 & \tilde{A} \end{bmatrix}, \quad
\bar{B} = \begin{bmatrix} B \\ \tilde{B} \end{bmatrix} \\
\bar{C} &= \begin{bmatrix} C & -\tilde{C}\end{bmatrix}, \quad \bar{D} = D - \tilde{D}.
\end{align}
Note that we skipped the $\theta$ dependence for simplicity of notation. Unless otherwise stated, all system matrices are assumed to be dependent on the parameters $\theta$. Now, we can write the augmented system dynamics as
\begin{nalign}
\label{eq:augmented_sys}
\dot{\bar{x}} &= \bar{A}\bar{x} + \bar{B}u \\
\bar{y} &= \bar{C}\bar{x} + \bar{D}u.
\end{nalign}
In this formulation, the output of the system $\bar{y}$ is the estimation error 
\begin{align}
\bar{y} = y - \tilde{y} = Cx - \tilde{C}\tilde{x} + Du - \tilde{D}u = \bar{C}\bar{x} + \bar{D}\bar{u}
\end{align}
Note that the total number of states in the augmented system is $2n$, the number of inputs is $k$, we have $m$ outputs and $\bar{D} \in \mathbb{R}^{m \times k}$, similar to $D$ and $\tilde{D}$. 
% $x$ and $\tilde{x}$ are both $n \times 1$, but estimated state values will differ because $\tilde{A}$ differs. Thus, we have a new dynamical system:
% \begin{align}
% \dot{\tilde{x}} &= \tilde{A}\tilde{x} + \tilde{B}u, \quad \tilde{y} = \tilde{C}\tilde{x} + \tilde{D}u
% \end{align}
Here, $u$ remains the same because $u$ represents physical inputs for a system with forced response.
\subsection{Parametric robustness of estimation error}
Using equation~\eqref{robustness_metric}, we can re-write the robustness metric $R$ using $\text{err} = \bar{y}$ as 
\begin{equation}
    \label{robustness_metric_est}
    R = \left(1 + \mathlarger{\mathlarger{\sum}}_{i = 1}^{p}\frac{\theta_i^*\norm{\left.\frac{\partial \bar{y}}{\partial \theta_i}\right\rvert_{\theta_i = \theta_i^*}}}{\norm{\bar{y}(t, \theta_i^*)}}\right)^{-1}.
\end{equation}
Thus, our goal is to compute a bound on $\norm{\partial \bar{y}/\partial \theta_i}$ for all parameters $\theta_i$. 
% To develop this metric, we discuss the main lemmas and existing results related to parametric sensitivities in the next section. 

%% file: sections/preliminaries.tex
\label{prelims}
\textbf{Notation: }In this paper, we consider the Euclidean 2-norm for vectors, $\norm{x}_2$ for the 2-norm of $x \in \mathbb{R}^n$. For matrices, $\norm{\cdot}$ represents the induced 2-norm. For all norms without the suffix, we assume the 2-norm. The $l^{\infty}$-norm for vectors is defined as the sup-norm, in other words, this norm is used to compute the maximum absolute value of the vector. For matrices that describe system dynamics, we use the logarithm norm using the lemma that follows.

\textbf{Assumptions:} We assume that both the original and the estimated system dynamics are exponentially stable dynamical systems. This ensures that the estimation error is not unbounded, that is, we assume the correctness of the estimation algorithm. Thus, our method can be applied to any data-driven estimation algorithm to compare the robustness of the methods given that the methods all lead to bounded estimation errors and that the system descriptions can be re-formulated in the structure required here. We also assume that $D = 0$, without loss of generality as it will add an additional term in the bound, which is easy to compute using $\norm{\bar{D}u}_{\infty}$. Finally, we assume that the systems considered in this paper are input-to-state (ISS) stable that allows us to use the $l^{\infty}$-norm bounds on the control input while developing the robustness metric.  

\begin{lemma}[See~\cite{matrix_exp_bound_proof_log_norm}]
\label{exponential_bound}
For a matrix $A$, the norm of the matrix exponential ($e^{At}$) is bounded as 
\begin{equation*}
    \norm{e^{At}} \leq e^{-\abs{\mu} t},
\end{equation*}
where $\mu$ is the logarithm norm of $A$~\cite{lognorm}, under the assumption that $A$ is Hurwitz. For the log-norm induced by the 2-norm, we have
\begin{equation*}
    \mu(A) = \frac{\lambda_{\max}(A + A^T)}{2},
\end{equation*}
Note that for Hurwitz $A$, the matrix exponential is decaying and thus, $\mu$ is always negative.
\end{lemma}
For partial parameter derivatives, the two lemmas that will be used in the derivation of the main result in the paper are:
\begin{lemma}
\label{lemma_matrix_exp}
For Hurwitz A, the partial derivative of $e^{At}$ with respect to a parameter $\theta_i \in \theta$ is bounded above as
\begin{equation}
\label{bound_derivative}
    \norm{\frac{\partial e^{At}}{\partial \theta_i}} \leq  \norm{\frac{\partial A}{\partial \theta_i}} te^{-\abs{\mu} t}.
\end{equation}
where $\abs{\mu}$ is the absolute value of the log-norm of $A$.
% as in Lemma~\ref{exponential_bound}.
\begin{proof}
Using the solution for linear dynamical system $x(t) = e^{At}x(0)$, we can write
\begin{equation*}
    \frac{\partial x(t)}{\partial \theta_i} = e^{At}\frac{\partial x(0)}{\partial \theta_i} + \frac{\partial e^{At}}{\partial \theta_i} x(0).
\end{equation*}
Then, using the convolution equation, the partial derivative of the matrix exponential is
\begin{equation}
\label{matrix_exp_derivative}
    \frac{\partial e^{At}}{\partial \theta_i} = \int_0^t e^{(t- \tau)A}\frac{\partial A}{\partial \theta_i}e^{\tau A} d\tau,
\end{equation}
which leads to the upper bound in the lemma on solving the integral after using Lemma~\ref{exponential_bound}.
\end{proof}
\end{lemma}

\begin{lemma}
\label{lemma_matrix_exp_negative}
The derivative of the negative matrix exponential \( e^{-As} \) with parameter \( \theta_i \) satisfies the following bound:
\begin{equation}
\label{eq:lemma_bound}
\left\|\frac{\partial e^{-As}}{\partial \theta_i}\right\| \leq \left\|\frac{\partial A}{\partial \theta_i}\right\| s e^{|\mu| s},
\end{equation}
where \( |\mu| \) is the absolute value of the log-norm of \(A\).
\end{lemma}

\begin{proof}
Using equation~(\ref{matrix_exp_derivative}) and substituting the negative time in the integral we obtain
\begin{equation*}
\frac{\partial e^{-As}}{\partial \theta_i} = \int_0^{-s} e^{A(-s-\tau)}\frac{\partial A}{\partial \theta_i}e^{A\tau} d\tau.
\end{equation*}
Using submultiplicativity, the substitution that $u = -\tau$, and Lemma~\ref{exponential_bound}, we get:
\begin{equation*}
\left\|\frac{\partial e^{-As}}{\partial \theta_i}\right\| \leq \left\|\frac{\partial A}{\partial \theta_i}\right\| e^{|\mu|s}\int_0^s du,
\end{equation*}
which leads to the desired result on solving the integral.
\end{proof}

%% file: sections/thm-A.tex
\label{results}
The main result of this paper is an upper bound on the robustness of the estimation error with respect to a model parameter $\theta_i$. This bound enables the quantification of the robustness metric defined in equation~(\ref{robustness_metric_est}) by computing the bound for each model parameter. The advantage of this approach is that the system modeler can individually evaluate the impact of physical parameters of interest and, accordingly, update the robustness computation. For example, if only a subset of parameters are of interest, then the robustness metric can be computed only for these parameters. 

We present the result assuming parametric uncertainties in $A$ and in the initial conditions $x(0)$ for simplicity because this is the most common case that would appear in estimating dynamics of physical systems --- the external inputs and output measurements are usually parameter independent. However, if the matrices $B$ and $C$ are also parameter dependent, the result can be extended.
% , as it will add more terms that quantify the parametric robustness contributions of the $B$ and $C$ matrices that are parameter dependent. 
\subsection{Parameter dependent A}
\begin{theorem}
For a parametrized estimated system with control inputs, the parametric robustness of the error in estimation for a given parameter $\theta_i$ is bounded above using a computable contribution of three terms: (1) the simple derivative of $\bar{A}$ with $\theta_i$, (2) the worst-case norm of inputs, and (3) the maximum time for which the inputs are applied:
\begin{nalign}
    \norm{\frac{\partial \bar{y}}{\partial \theta_i}}^2 \leq K_1 \norm{\frac{\partial \bar{A}}{\partial \theta_i}}^2 &+ K_2 \norm{\frac{\partial \bar{A}}{\partial \theta_i}}^3 \norm{\bar{B}u}_{\infty} \\ &+ K_3 N^2\norm{\frac{\partial \bar{A}}{\partial \theta_i}}^2\norm{\bar{B}u}_{\infty}, \label{thm:A}
\end{nalign}
where $K_1, K_2,$ and $K_3$ are positive constants dependent on the initial condition, the log-norm of $\bar{A}$, the augmented output matrix $\bar{C}$, and the maximum time horizon $N$.
\end{theorem}
\begin{proof}
The main proof idea is to describe the parametric robustness of the estimation error in the form of parametric robustness of the matrix exponential, which enables the use of the lemmas developed earlier. We start by using the convolution equation~\cite{aastrom2021feedback}. Write \(\bar{y}(t)\) for the augmented system as:
\begin{align}
\label{eq:sys_soln}
\bar{y}(t) &= \bar{C}e^{\bar{A}t}\bar{x}(0) + \int_0^t \bar{C}e^{\bar{A}(t-\tau)}\bar{B}u(\tau)d\tau.
\end{align} 
Then, using the definition of the Euclidean norm and the equation above, we can write the left-hand side (LHS) of the theorem 
(the square of 2-norm of $\partial \bar{y}/\partial \theta_i$) 
as
\begin{align*}
\int_0^{\infty}\frac{\partial}{\partial \theta_i} \left[\int_0^t u^T(\tau)\bar{B}^Te^{\bar{A}^T(t-\tau)}\bar{C}^T d\tau + \bar{x}^T(0)e^{\bar{A}^T t}\bar{C}^T \right] \\ \frac{\partial}{\partial \theta_i}\left[\bar{C}e^{\bar{A}t}\bar{x}(0) + \int_0^t \bar{C}e^{\bar{A}(t-\tau)}\bar{B}u(\tau)d\tau\right] dt
\end{align*}
Expanding further, we obtain:
\begin{align*}
&= \int_0^{\infty}\underbrace{\left[\int_0^t  u^T(\tau)\bar{B}^T\frac{\partial e^{\bar{A}^T(t-\tau)}}{\partial \theta_i}\bar{C}^T d\tau + \bar{x}^T(0)\frac{\partial e^{\bar{A}^T t}}{\partial \theta_i}\bar{C}^T\right]}_{(a)} \\
& \underbrace{\left[ \bar{C}\frac{\partial e^{\bar{A}t}}{\partial \theta}\bar{x}(0) + \int_0^t \bar{C}\frac{\partial e^{\bar{A}(t-\tau)}}{\partial \theta}\bar{B}u(\tau)d\tau \right]}_{(b)} dt
\end{align*}
We break down the above integral into four parts that result from the multiplication of the two additive terms in each of (a) and (b) above: $a_1b_1, a_1b_2, a_2b_1, a_2b_2$. Since $\norm{a_1b_1} = \norm{a_2b_2}$, we club these together later and start with $a_1b_2$:
\begin{align*}
% a_1b_1 = \int_0^\infty &\left(\int_0^t u^T(\tau)\bar{B}^T\frac{\partial e^{\bar{A}^T(t-\tau)}}{\partial \theta_i}\bar{C}^T d\tau \right) \\ 
% &\left(\bar{C}\frac{\partial e^{\bar{A}^T t}}{\partial \theta_i}\bar{x}(0)\right) dt \\
a_1b_2 = \int_0^\infty &\left(\int_0^t u^T(\tau)\bar{B}^T\frac{\partial e^{\bar{A}^T(t-\tau)}}{\partial \theta_i}\bar{C}^T d\tau\right)
\\
&\left(\int_0^t \bar{C}\frac{\partial e^{\bar{A}(t-\tau)}}{\partial \theta_i}\bar{B}u(\tau)d\tau\right) dt \end{align*}
which is equal to the 2-norm by definition,
\begin{align*}
a_1b_2 &= \left\|\int_0^t \bar{C}\frac{\partial e^{\bar{A}(t-\tau)}}{\partial \theta_i}\bar{B}u(\tau)d\tau\right\|_2^2.
\end{align*}
For this, we expand the partial derivative of the matrix exponential such that we can use Lemma~\ref{lemma_matrix_exp_negative}, to obtain
\begin{align*}
a_{1}b_{2} &\leq \|\bar{C}\|\left\|\frac{\partial \bar{A}}{\partial \theta_i}\right\|^2\|\bar{B}u\|_{\infty}\left[\frac{t^2 \abs{\mu} + te^{-|\mu|t} - t}{|\mu|^2}\right]
\end{align*}
by solving the integral of $te^{\abs{\mu}t}$ (the residual term in Lemma~\ref{lemma_matrix_exp_negative}) from $0$ to $\infty$. Note that for all $t$, since $te^{-\abs{\mu}t}$ is always less than $t$ and we integrate until a maximum time of $N$, we can write a conservative upper bound,
\begin{align}
\label{eq:a1b2}
a_{1}b_{2} &\leq \frac{N^2}{|\mu|}\|\bar{C}\|\left\|\frac{\partial \bar{A}}{\partial \theta_i}\right\|^2\|\bar{B}u\|_{\infty}.
\end{align}
Then, for the third term $a_2b_1$, we write
\begin{align*}
a_2b_1 &= \int_0^{\infty} \bar{x}(0)^T \frac{\partial e^{\bar{A}^Tt}}{\partial \theta_i}\bar{C}^TC\frac{\partial e^{\bar{A}t}}{\partial \theta_i}\bar{x}(0)dt.
\end{align*}
Using the sub-multiplicative property of the 2-norm we get,
\begin{equation*}
 a_2b_1 \leq \int_0^{\infty} \norm{\frac{\partial e^{\bar{A}t}}{\partial \theta_i}}^2\norm{\bar{C}^T\bar{C}}\norm{\bar{x}(0)}^2 dt,
\end{equation*}
then, using Lemma~\ref{lemma_matrix_exp} and evaluating the integral
% \begin{align*}
    % \int_N^{\infty} t^2 e^{-2\abs{\mu} t}dt = \frac{N^2e^{-2\abs{\mu} N}}{2\abs{\mu}} + \frac{Ne^{-2\abs{\mu} N}}{2\abs{\mu}^2} + \frac{e^{-2\abs{\mu} N}}{4\abs{\mu}^3}. 
    $\int_0^{\infty} t^2 e^{-2\abs{\mu} t}dt =1/(4\abs{\mu}^3)$
% \end{align*}
we finally get that
\begin{align}
\label{eq:a2b1}
a_2b_1 &\leq \frac{1}{4\abs{\mu}^3}\left\|\frac{\partial \bar{A}}{\partial \theta_i}\right\|_2^2 \left\|\bar{C}^T\bar{C}\right\|_2 \left\|\bar{x}(0)\right\|_2^2 
\end{align}
% where $\abs{\mu}$ is the absolute value of the log-norm of $\bar{A}$. Then,  
We expand the last term, $a_2b_2$ in Appendix A to get
% \begin{align*}
% a_2b_2 &= \int_0^\infty \bar{x}^T(0)\frac{\partial e^{\bar{A}^T t}}{\partial \theta_i}\bar{C}^T \left(\int_0^t \bar{C}\frac{\partial e^{\bar{A}(t-\tau)}}{\partial \theta_i}\bar{B}u(\tau)d\tau\right) dt.
% \end{align*}
% Note that the norm for this term is the same as $a_1b_1$. The full expansion and bound for this term is shown in Appendix A. We give the final bound for these two terms together:
% Summarizing and concluding for a_2b_2 (First image)
\begin{align}
\label{eq:a2b2}
a_{2}b_{2} + a_1b_1 &\leq \frac{2}{|\mu|^5}\left(\|\bar{x}(0)\|\|\bar{C}^T\bar{C}\| \left\|\frac{\partial \bar{A}}{\partial \theta_i}\right\|^3 \|\bar{B}u\|_{\infty}\right)
\end{align}
Adding equations~(\ref{eq:a1b2},~\ref{eq:a2b1},~\ref{eq:a2b2}) we obtain the desired result with $K_3 = \|\bar{C}\|/|\mu|$, and $K_1, K_2$ as
\begin{align*}
    K_1  = \frac{\left\|\bar{C}^T\bar{C}\right\|_2 \left\|\bar{x}(0)\right\|_2^2}{4\abs{\mu}^3}, \: \:
    K_2 = \frac{2\|\bar{x}(0)\|\|\bar{C}^T\bar{C}\|}{|\mu|^5}.\: \quad 
    \qedhere
\end{align*}
% \begin{align}
% &a_{1}b_{1}+a_{1}b_{2}+a_{2}b_{1}+a_{2}b_{2} \nonumber \\
% &\quad\leq \frac{2\|\bar{x}(0)\|\|\bar{C}^T\bar{C}\| \left\|\frac{\partial \bar{A}}{\partial \theta_i}\right\|^3 \|\bar{B}u\|_{\infty}}{|\mu|^5}
% \\&+ \frac{\|\bar{C}\|\left\|\frac{\partial \bar{A}}{\partial \theta_i}\right\|^2\|\bar{B}u\|_{\infty}}{|\mu|^2}\left[\frac{\mu t e^{|\mu|t}-e^{|\mu|t}+1}{|\mu|^2}\right]
% \\&+\tilde{M}\left\|\frac{\partial \bar{A}}{\partial \theta_i}\right\|^2\|\bar{C}^T\bar{C}\|_2\|\bar{x}(0)\|^2_2
% \end{align}
\end{proof}
\subsection{Parameter dependent initial conditions}
Here, we assume that initial conditions are dependent on parameters and the dynamics are parameter independent. 
% Then, we can obtain a simpler bound on the parametric robustness of the estimation error as
\begin{theorem}
    For a given estimation of a parameter independent, observable, linear dynamical system with control inputs and parameter dependent initial conditions, the parametric robustness of the error in estimation for a parameter $\theta_i$ is bounded above by
\begin{align}
   \norm{\frac{\partial \bar y}{\partial \theta_i}}^2 \leq \lambda_{\max}(P)\norm{\frac{\partial\bar{x}(0)}{\partial \theta_i}}^2 \label{thm:x0},
\end{align}
where $P$ is the Lyapunov matrix that satisfies the observability equation $\bar{A}^TP + P\bar{A} = -\bar{C}^T\bar{C}$.
\begin{proof}
Similar to the previous proof, we find that all terms except the one dependent on the sensitivity of the initial condition will go to 0. This is because the system dynamics are parameter independent and on integrating the matrix exponential factors from $0$ to $\infty$, bounded above by Lemma~\ref{lemma_matrix_exp}, will asymptotically go to zero, leaving only the following term
\begin{align*}
        \norm{\frac{\partial \bar{y}}{\partial \theta_i}}^2 \leq \int_0^{\infty} \frac{\partial \bar{x}(0)^T}{\partial \theta_i} e^{\bar{A}^Tt} \bar{C}^T\bar{C}e^{\bar{A}t}\frac{\partial \bar{x}(0)}{\partial \theta_i} dt.
\end{align*}
The rest of the proof follows the proof method in~\cite{pandey2023robustness}.
% where it is applied to the model reduction robustness problem. 
Specifically, we have from~\cite[Ch.5]{antsaklis}, that for an asymptotically stable system there exists a unique matrix $P$ that satisfies the observability Gramian:
\begin{align*}
    P &= \lim_{N\to\infty} W_{\text{o}}(N) = \lim_{N\to\infty}\int_{0}^{N}e^{\bar{A}^Tt}\bar{C}^T\bar{C}e^{\bar{A}t}dt,
\end{align*}
% where $W_{\text{o}}(N)$ is the observability Gramian. 
Substituting this above and bounding by the maximum eigen value of $P$ gives us the desired result.
\end{proof}
\end{theorem}

%% file: sections/results.tex
\subsection{Special Cases}
To validate our results, we study the following special cases
\textbf{Case 1:} \(u = 0\) (No forced input)
\begin{align}
a_{1}b_{1} = a_{1}b_{2} = a_{2}b_{2} = 0, \quad \text{thus,}
\end{align}
\begin{align}
a_{2}b_{1} &\leq \frac{1}{4\abs{\mu}^3}\left\|\frac{\partial \bar{A}}{\partial \theta_i}\right\|^2\|\bar{C}^T\bar{C}\|_2\|\bar{x}(0)\|^2_2,
\end{align}
which is the same result as obtained in a previously published article~\cite{pandey2021robustness} on the parametric robustness of model reduction. Note that the main result of this paper is to extend these proof techniques to systems with forced inputs, and for data-driven system estimation problems.

\textbf{Case 2:} \(\bar{x}(0) = 0\) (Only forced response)

We have $a_{2}b_{1}= a_{1}b_{1}= a_{2}b_{2} = 0$, thus, we are left with
\begin{align}
\norm{\frac{\partial \bar{y}}{\partial \theta_i}}^2 \leq K_3 N^2\norm{\frac{\partial \bar{A}}{\partial \theta_i}}^2\norm{\bar{B}u}_{\infty}.
\end{align}
The equivalence of the results in this paper with other findings validate our proposed bounds. 
% \subsection{Equivalence With Other Methods}
% \subsection{Algorithm?}

%% file: sections/applications.tex
\label{applications}
Now we will apply the methodology outlined above to compute the robustness upper bound for estimation of parametric linear dynamical systems. By computing the upper bound, we aim to show the effectiveness of the metric in capturing the robustness level for a given parametric uncertainty on parameters of interest. The robustness metric gives the worst case guarantee on the changes to the estimation error, as opposed to running the estimation algorithm multiple times for different parameter samples. 
\begin{figure}
    \centering
    \includegraphics[scale=0.5]{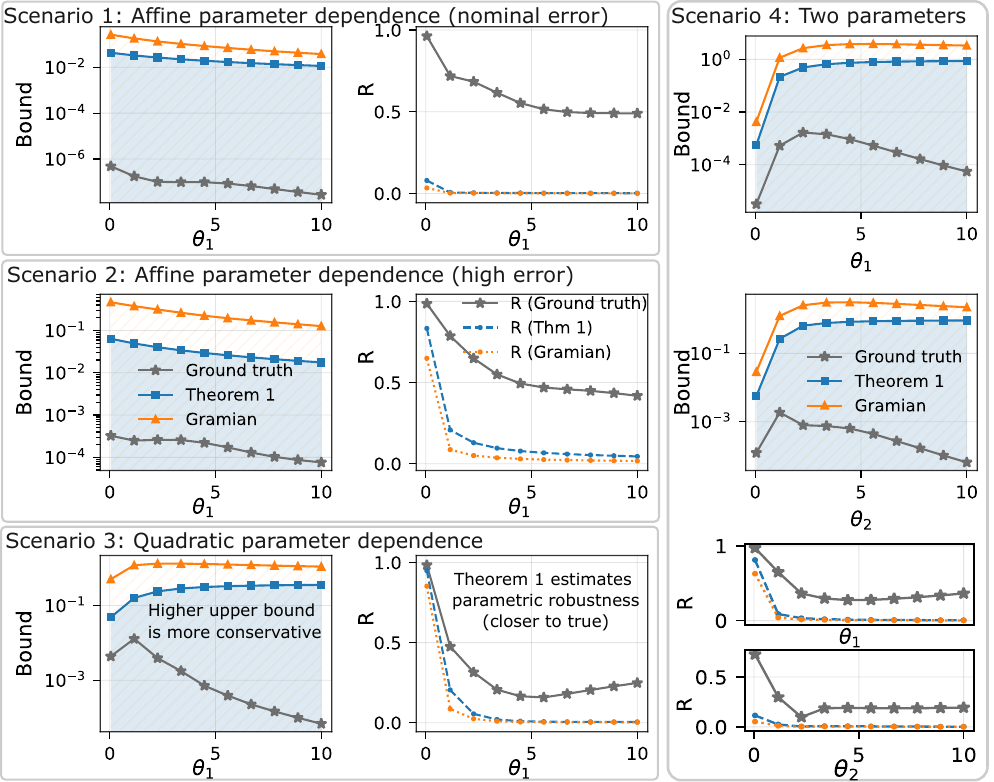}
    \caption{\footnotesize Sensitivity of estimation error in four different scenarios for a linear parameter varying $A$. Note that a lower value of the upper bound is less conservative, which corresponds to a higher $R$.}
    \label{fig:results}
\end{figure}
\subsection{Parametric uncertainty in $A$}
Consider a mass-spring damper system with a scheduling parameter $\theta_1$. The states ($x$) of a simplified 2D dynamics are the position ($q$) and velocity ($\dot{q}$) of the mass object. The dynamics are given by:  
$ \dot{x} = A(\theta)x + Bu $, and $ y = Cx + Du $
where 
\begin{align*}
    A(\theta) = \begin{bmatrix} 0 & 1 \\ -(20 + 5\theta_1) & -(2 + 0.5\theta_1) \end{bmatrix} \quad 
    B = \begin{bmatrix} 0 \\ 1 \end{bmatrix},
\end{align*}
and $C = \begin{bmatrix} 1 & 0 \end{bmatrix}, D = 0$. Then, we can write $A(\theta)$ in an affine form as: $ A(\theta) = A_0 + A_1\theta$
with
\begin{align*}
    A_0 = \begin{bmatrix} 0 & 1 \\ -20 & -2 \end{bmatrix} , \quad  A_1 = \begin{bmatrix} 0 & 0 \\ -5 & -0.5 \end{bmatrix}
\end{align*}
We assume an estimated system dynamics
\begin{align*}
   \tilde{A} = \begin{bmatrix}0 & 1 \\ -(19.80 + 5.10\theta_1) & -(2.05 + 0.48\theta_1)\end{bmatrix},
\end{align*}
which has a clear affine form. Now, to apply Theorem 1, we can write the augmented dynamics of the $\bar{x}$ system.
% {\footnotesize
% \begin{align*}
%     \bar{A} = \begin{bmatrix} 0 & 1 & 0 & 0\\ -(20 + 5\theta_1) & -(2 + 0.5\theta_1) & 0 & 0 \\ 0 & 0 & 0 & 1 \\0 & 0 &  -(19.80 + 5.10\theta_1) & -(2.05 + 0.48\theta_1)\end{bmatrix}
% \end{align*}}
Then, the derivative of $\bar{A}$ with respect to $\theta_1$ can be computed symbolically. We introduce four different scenarios to measure (and compare) the parameter-dependent sensitivity of the estimation error: (1) Affine dependence on parameter $\theta_1$, (2) Affine dependence with inaccurate parameter identification by changing the constants in $\tilde{A}$, (3) Quadratic dependence, in addition to affine, on $\theta_1$, (4) Dependence on two parameters $\theta_1$ and $\theta_2$, along with a joint term $\theta_1 \theta_2$. Details of the four scenarios are available in the associated \href{https://github.com/ayush9pandey/robest}{\underline{paper code}}
and the results are shown in Figure~\ref{fig:results} for the bound given by Theorem 1 and the robustness metric $R$ given in equation~(\ref{robustness_metric}). For comparison, we use a bound based on the trace of the observability Gramian. See Appendix B for derivation of this baseline comparison bound for the sensitivity of the estimation error in LPV systems. We note that the bound given by Theorem 1, which actively exploits the stability of the augmented system, leads to less conservatism than the Gramian-based bound, which is more general. 
\subsection{Parametric uncertainty in $x(0)$}
\begin{figure}
    \centering
    \includegraphics[width=\linewidth]{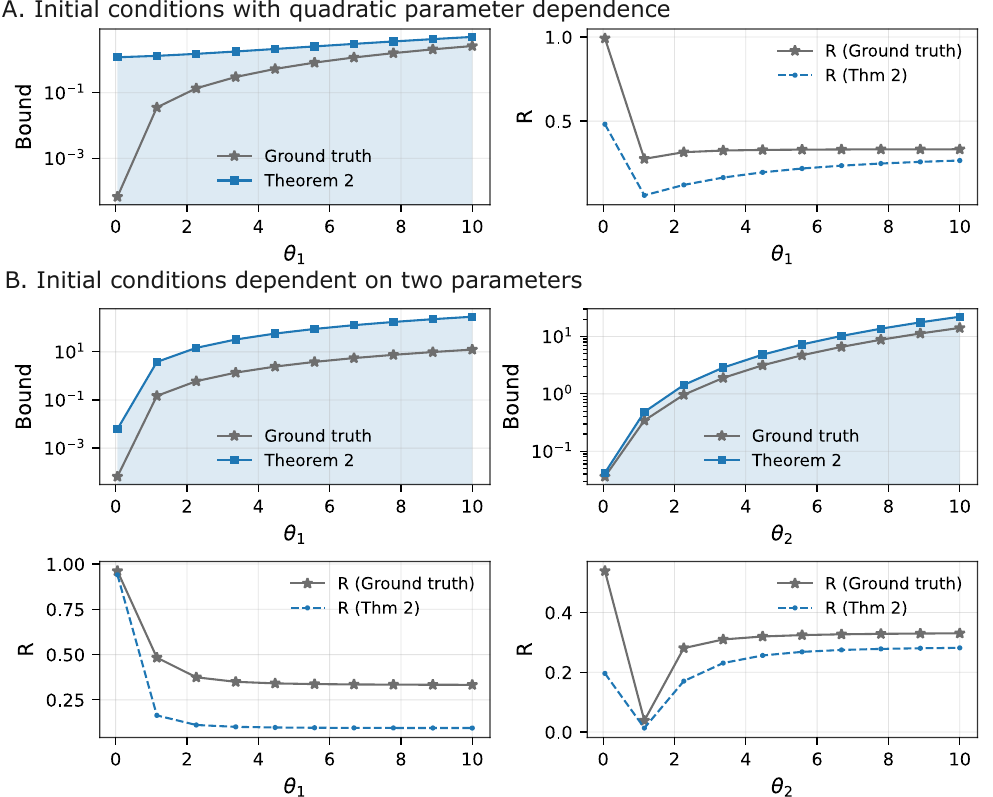}
    \caption{\footnotesize Sensitivity of estimation error when the initial conditions are parameter dependent (bound using Theorem 2).}
    \label{fig:results_thm2}
\end{figure}

For parameter-dependent initial conditions, we apply Theorem 2 to compute the upper bound and compare this with the ground truth computation of $\partial \bar y/ \partial \theta$ by solving the system trajectories. We show the upper bound for two different scenarios where parameter dependence of $x(0)$ is significant -- quadratic dependence (see Figure~\ref{fig:results_thm2}A) and dependence on two parameters (see Figure~\ref{fig:results_thm2}B)
\subsection{Computational complexity}
To compute the ground truth sensitivity of the estimation error $\norm{\partial \bar{y}/\partial \theta}$, the system must be simulated four times if the central difference method is used for derivative computation (twice each for original system and the estimated system). For accurate derivative computation, more number of simulations may be required. Therefore, the computation of the ground truth is computationally intensive. In comparison, the proposed bounds in this paper can be computed using only the symbolic derivative of $A$, without solving for the full system dynamics --- a significant advantage as the system model grows bigger.   

%% file: sections/conclusions.tex
% \label{conclusions}
In this paper, we consider the problem of parametric dependence in linear dynamical systems. We introduce a new method to compute the robustness of estimation error for linear dynamical systems with control inputs. This robustness metric is motivated by a sensitivity analysis perspective, quantifying the rate of change of error with respect to parametric uncertainties in the dynamics or the initial conditions. The main contribution is a closed-form upper bound for the robustness metric that can be selectively computed for parameters of interest without solving for the system trajectories. This approach enables systematic comparisons of various data-driven estimation methods in parametric dynamical systems, aiding researchers in selecting effective estimation strategies or combinations thereof. 
% Future research directions include an extensive exploration of application case studies to further assess the performance of the proposed metric across different estimation methods.

%% file: sections/appendix_a2b2.tex
\section*{Appendix A}
For the $a_2b_2$ term in the proof of Theorem 1, we can obtain the bound as follows.
{\footnotesize
\begin{align*}
a_{2}b_{2} = \int_0^\infty &\bar{x}^T(0)\frac{\partial e^{\bar{A}^T t}}{\partial \theta}\bar{C}^T\left(\int_0^t\bar{C}\frac{\partial e^{\bar{A}(t-\tau)}}{\partial \theta}\bar{B}u(\tau)d\tau\right) dt \\
\leq \int_0^\infty &\|\bar{x}(0)\| \left\|\frac{\partial e^{\bar{A}^T t}}{\partial \theta}\right\|^2 \|\bar{C}^T\bar{C}\| \\&\left\|\int_0^t \frac{\partial e^{-\bar{A}\tau}}{\partial \theta}\bar{B}u(\tau)d\tau\right\| dt \\
\leq &\|\bar{x}(0)\|\|\bar{C}^T\bar{C}\|\left\|\frac{\partial \bar{A}}{\partial \theta}\right\|^2\\ &\int_0^\infty t^2 e^{-2|\mu|t}\left\|\int_0^t \frac{\partial e^{-\bar{A}\tau}}{\partial \theta}\bar{B}u(\tau)d\tau\right\| dt
\end{align*}
}
\textbf{Aside:} To bound the norm of the integral on the right, we use Lemma~\ref{lemma_matrix_exp} to write
{\footnotesize \begin{align*}
\left\|\int_0^t \frac{\partial e^{-\bar{A}\tau}}{\partial \theta}\bar{B}u(\tau)d\tau\right\| &\leq \|\bar{B}u\|_{\infty}\int_0^t\left\|\frac{\partial e^{-\bar{A}\tau}}{\partial \theta_i}\right\| d\tau \\
&\leq \|\bar{B}u\|_{\infty}\int_0^t\left\|\frac{\partial \bar{A}}{\partial \theta_i}\right\|\tau e^{|\mu|\tau} d\tau \\
\leq \|\bar{B}u\|_{\infty}&\left\|\frac{\partial \bar{A}}{\partial \theta_i}\right\|\frac{[e^{|\mu|t}(|\mu|t - 1) + 1]}{|\mu|^2}
\end{align*}}
This result can now be substituted back into the above integral to complete the bound on \(a_{2}b_{2}\):
{\footnotesize\begin{align*}
a_{2}b_{2} \leq &\|\bar{x}(0)\|\|\bar{C}^T\bar{C}\|\left\|\frac{\partial \bar{A}}{\partial \theta_i}\right\|^2 \times\\ &\int_0^\infty t^2 e^{-2|\mu|t}\left\|\int_0^t\frac{\partial e^{-\bar{A}\tau}}{\partial \theta}\bar{B}u(\tau)d\tau\right\| dt \\
&\leq \|\bar{x}(0)\|\|\bar{C}^T\bar{C}\|\left\|\frac{\partial \bar{A}}{\partial \theta}\right\|^2 \|\bar{B}u\|_{\infty}\left\|\frac{\partial \bar{A}}{\partial \theta_i}\right\| \frac{1}{|\mu|^2} I_1
\\
\text{where}\\
I_1 &= \int_0^\infty t^2 e^{-2|\mu|t}\left[e^{|\mu|t}(|\mu|t - 1) + 1\right] dt
\end{align*}}

Breaking the integral down into three separate integrals, we have:
{\footnotesize\begin{align*}
&= \frac{\|\bar{x}(0)\|\|\bar{C}^T\bar{C}\|\left\|\frac{\partial \bar{A}}{\partial \theta}\right\|^3\|\bar{B}u\|_{\infty}}{|\mu|}\int_0^\infty t^3 e^{-|\mu|t} dt \\
&\quad -\frac{\|\bar{x}(0)\|\|\bar{C}^T\bar{C}\|\left\|\frac{\partial \bar{A}}{\partial \theta}\right\|^3\|\bar{B}u\|_{\infty}}{|\mu|^2}\int_0^\infty t^2 e^{-|\mu|t} dt \\
&\quad +\frac{\|\bar{x}(0)\|\|\bar{C}^T\bar{C}\|\left\|\frac{\partial \bar{A}}{\partial \theta}\right\|^3\|\bar{B}u\|_{\infty}}{|\mu|^2}\int_0^\infty t^2 e^{-2|\mu|t} dt
\end{align*}}
These three integrals can then be individually evaluated with limits $0$ to $\infty$ as: \(\int t^3 e^{-|\mu| t}\, dt = 6/|\mu|^4\); \(\int t^2 e^{-|\mu| t}\, dt = 2/|\mu|^3\); \(\int t^2 e^{-2|\mu| t}\, dt = 1/(4|\mu|^3)\). Substituting these integrals, we get the desired result used in the proof.

%% file: sections/appendix_baseline.tex
\section*{Appendix B}
\subsection*{Baseline upper bound for the sensitivity of estimation error}
\begin{theorem}
In the finite time horizone $[0, N]$, the $L^2$ norm of sensitivity of the estimation error is bounded above
\begin{equation}
\left\|\frac{\partial \bar{y}}{\partial \theta_i}\right\|_{2,[0,N]}^2 \;\le\; N\,\mathrm{trace}\!\big(\bar Q_o(N)\big)\;\|\bar w\|_{2,[0,N]}^2\;
\end{equation}
where $Q_o(N)$ is the finite time horizon observability Gramian and $w$ models the system trajectories $(\partial \bar{A}/\partial \theta_i)\bar{x}$.
\end{theorem}
% \subsection*{A baseline upper bound for sensitivity of the estimation error}
% In this Appendix, we derive an upper bound on $\bar{y}_s := \partial \bar{y}/\partial \theta_i$ by using system trajectories $\bar{x}$ explicitly.
\begin{proof}
Write $\bar{z} = \partial \bar{x}/\partial \theta_i$ as the sensitivity coefficient of the augmented system given in equation~\eqref{eq:augmented_sys}. Then, the sensitivity system equation is $
\dot{\bar z}(t)=\bar A\,\bar z(t)+\bar w(t),\: \bar z(0)=0,\:
\partial \bar{y}/ \partial \theta_i = \bar C\,\bar z.$
Using the convolution equation~(\ref{eq:sys_soln}) for the sensitivity system, we can write the $L^2$ norm on a finite horizon $[0,N]$
% \[
% y_s(t)=\int_{0}^{t} \bar C\,e^{\bar A (t-\tau)}\,\bar w(\tau)\,d\tau .
% \]
% We are interested in bounding the induced L2 norm of the above, computed on a finite horizon $[0, N]$. 
\begin{align*}
&\left\|\frac{\partial \bar y}{\partial \theta_i}\right\|_{2,[0,N]}^2=\int_{0}^{N}\!\!\left\|\frac{\partial \bar y}{\partial \theta_i} (t)\right\|_2^2\,dt \\ 
&\;\le\;\int_{0}^{N}\!\!\Big(\int_{0}^{t}\!\!\big\|\bar C e^{\bar A (t-\tau)}\big\|_F^2 d\tau\Big)
\Big(\int_{0}^{t}\!\!\|\bar w(\tau)\|_2^2 d\tau\Big)\,dt.
\end{align*}
where the bound is obtained using Cauchy-Schwarz inequality for each $t$. 
% Define the matrix $\bar K(t,\tau):=\bar C e^{\bar A (t-\tau)}\mathbf{1}_{\{\tau\le t\}}$. 
% The inequality above can be then represented as the standard operator bound $\| \bar K \|_{2\to 2}\le \|\bar K\|_{\mathrm{HS}}$, where for matrix kernels the Hilbert–Schmidt norm satisfies 
% $\|\bar K\|_{\mathrm{HS}}^2=\int_0^N\!\!\int_0^N \|\bar K(t,\tau)\|_F^2\,d\tau\,dt$.
Using the finite-time horizon, $\int_0^{t}\|\bar w(\tau)\|_2^2 d\tau \le \|\bar w\|_{2,[0,N]}^2$ and the change of variables $s=t-\tau$ gives
\begin{align*}
\left\|\frac{\partial \bar y}{\partial \theta_i} \right\|_{2,[0,N]}^2
&\;\le\;\|\bar w\|_{2,[0,N]}^2
\int_{0}^{N}\!\!\int_{0}^{t}\!\!\big\|\bar C e^{\bar A (t-\tau)}\big\|_F^2 d\tau\,dt\\
&=\|\bar w\|_{2,[0,N]}^2\int_0^N (N-s)\,\|\bar C e^{\bar A s}\|_F^2 ds\\
& \le\; N\,\|\bar w\|_{2,[0,N]}^2\int_0^N \|\bar C e^{\bar A s}\|_F^2 ds.
\end{align*}

The Frobenius–trace identity $\|M\|_F^2=\mathrm{trace}(M^\top M)$ and trace cyclicity give us the final bound
\begin{align*}
&\int_0^N \|\bar C e^{\bar A s}\|_F^2 ds
=\int_0^N \mathrm{trace}\!\big(e^{\bar A^\top s}\bar C^\top\bar C\,e^{\bar A s}\big)ds\\
&=\mathrm{trace}\!\Big(\int_0^N e^{\bar A^\top s}\bar C^\top\bar C\,e^{\bar A s}ds\Big)
=\mathrm{trace}\big(\bar Q_o(N)\big),
\end{align*}
where $\bar Q_o(N)$ is the finite-horizon observability Gramian. Therefore,
\[
\left\|\frac{\partial \bar y}{\partial \theta_i}\right\|_{2,[0,N]}^2 \;\le\; N\,\mathrm{trace}\!\big(\bar Q_o(N)\big)\;\|\bar w\|_{2,[0,N]}^2.\;
\qedhere
\]
\end{proof}

%% file: main.bbl
\begin{thebibliography}{10}
\providecommand{\url}[1]{#1}
\csname url@samestyle\endcsname
\providecommand{\newblock}{\relax}
\providecommand{\bibinfo}[2]{#2}
\providecommand{\BIBentrySTDinterwordspacing}{\spaceskip=0pt\relax}
\providecommand{\BIBentryALTinterwordstretchfactor}{4}
\providecommand{\BIBentryALTinterwordspacing}{\spaceskip=\fontdimen2\font plus
\BIBentryALTinterwordstretchfactor\fontdimen3\font minus \fontdimen4\font\relax}
\providecommand{\BIBforeignlanguage}[2]{{%
\expandafter\ifx\csname l@#1\endcsname\relax
\typeout{** WARNING: IEEEtran.bst: No hyphenation pattern has been}%
\typeout{** loaded for the language `#1'. Using the pattern for}%
\typeout{** the default language instead.}%
\else
\language=\csname l@#1\endcsname
\fi
#2}}
\providecommand{\BIBdecl}{\relax}
\BIBdecl

\bibitem{schmid2010dynamic}
P.~J. Schmid, ``Dynamic mode decomposition of numerical and experimental data,'' \emph{Journal of Fluid Mechanics}, vol. 656, pp. 5--28, 2010.

\bibitem{proctor2016dynamic}
J.~L. Proctor, S.~L. Brunton, and J.~N. Kutz, ``Dynamic mode decomposition with control,'' \emph{SIAM Journal on Applied Dynamical Systems}, vol.~15, no.~1, pp. 142--161, 2016.

\bibitem{dawson2016characterizing}
S.~T. Dawson, M.~S. Hemati, M.~O. Williams, and C.~W. Rowley, ``Characterizing and correcting for the effect of sensor noise in the dynamic mode decomposition,'' \emph{Experiments in Fluids}, vol.~57, pp. 1--19, 2016.

\bibitem{demo2018pydmd}
N.~Demo, M.~Tezzele, and G.~Rozza, ``{PyDMD}: {P}ython dynamic mode decomposition,'' \emph{Journal of Open Source Software}, vol.~3, no.~22, p. 530, 2018.

\bibitem{ichinaga2024pydmd}
S.~M. Ichinaga, F.~Andreuzzi, N.~Demo, M.~Tezzele, K.~Lapo, G.~Rozza, S.~L. Brunton, and J.~N. Kutz, ``Pydmd: A python package for robust dynamic mode decomposition,'' \emph{Journal of Machine Learning Research}, vol.~25, no. 417, pp. 1--9, 2024.

\bibitem{brunton2016discovering}
S.~L. Brunton, J.~L. Proctor, and J.~N. Kutz, ``Discovering governing equations from data by sparse identification of nonlinear dynamical systems,'' \emph{Proceedings of the National Academy of Sciences}, vol. 113, no.~15, pp. 3932--3937, 2016.

\bibitem{brunton2021modern}
S.~L. Brunton, M.~Budi{\v{s}}i{\'c}, E.~Kaiser, and J.~N. Kutz, ``Modern koopman theory for dynamical systems,'' \emph{arXiv:2102.12086}, 2021.

\bibitem{darabi2025combining}
A.~Darabi, Z.~An, M.~A. Al-Radhawi, W.~Cho, M.~Siami, and E.~D. Sontag, ``Combining model-based and data-driven models: an application to synthetic biology resource competition,'' \emph{bioRxiv}, pp. 2025--03, 2025.

\bibitem{champion2019data}
K.~Champion, B.~Lusch, J.~N. Kutz, and S.~L. Brunton, ``Data-driven discovery of coordinates and governing equations,'' \emph{Proceedings of the National Academy of Sciences}, vol. 116, no.~45, pp. 22\,445--22\,451, 2019.

\bibitem{lee2024robot}
T.~Lee, J.~Kwon, P.~M. Wensing, and F.~C. Park, ``Robot model identification and learning: A modern perspective,'' \emph{Annual Review of Control, Robotics, and Autonomous Systems}, vol.~7, 2024.

\bibitem{martin2023guarantees}
T.~Martin, T.~B. Sch{\"o}n, and F.~Allg{\"o}wer, ``Guarantees for data-driven control of nonlinear systems using semidefinite programming: A survey,'' \emph{Annual Reviews in Control}, vol.~56, p. 100911, 2023.

\bibitem{pacti-paper}
I.~Incer, A.~Badithela, J.~B. Graebener, P.~Mallozzi, A.~Pandey, N.~Rouquette, S.-J. Yu, A.~Benveniste, B.~Caillaud, R.~M. Murray \emph{et~al.}, ``Pacti: Assume-guarantee contracts for efficient compositional analysis and design,'' \emph{ACM Transactions on Cyber-Physical Systems}, vol.~9, no.~1, pp. 1--35, 2025.

\bibitem{ames-guarantees}
T.~G. Molnar, R.~K. Cosner, A.~W. Singletary, W.~Ubellacker, and A.~D. Ames, ``Model-free safety-critical control for robotic systems,'' \emph{IEEE Robotics and Automation Letters}, vol.~7, no.~2, pp. 944--951, 2021.

\bibitem{chuchu-fan}
Y.~Meng, S.~Vemprela, R.~Bonatti, C.~Fan, and A.~Kapoor, ``Conbat: Control barrier transformer for safe robot learning from demonstrations,'' in \emph{2024 IEEE International Conference on Robotics and Automation (ICRA)}.\hskip 1em plus 0.5em minus 0.4em\relax IEEE, 2024, pp. 12\,857--12\,864.

\bibitem{nik-matni}
T.~T. Zhang, B.~D. Lee, I.~Ziemann, G.~J. Pappas, and N.~Matni, ``Guarantees for nonlinear representation learning: non-identical covariates, dependent data, fewer samples,'' \emph{arXiv:2410.11227}, 2024.

\bibitem{doyle2013feedback}
J.~C. Doyle, B.~A. Francis, and A.~R. Tannenbaum, \emph{Feedback Control Theory}.\hskip 1em plus 0.5em minus 0.4em\relax Courier Corporation, 2013.

\bibitem{gupta2023direct}
V.~Gupta, A.~Karimi, F.~Wildi, and J.-P. V{\'e}ran, ``Direct data-driven vibration control for adaptive optics,'' in \emph{2023 62nd IEEE Conference on Decision and Control (CDC)}.\hskip 1em plus 0.5em minus 0.4em\relax IEEE, 2023, pp. 8521--8526.

\bibitem{Schuchert2023}
P.~L. Schuchert, V.~Gupta, and A.~Karimi, ``Data-driven fixed-structure frequency-based $\mathbb{H}_2$ and $\mathbb{H}_{\infty}$ controller design,'' \emph{Automatica}, vol. 160, p. 110052, 2024.

\bibitem{baddoo2023physics}
P.~J. Baddoo, B.~Herrmann, B.~J. McKeon, J.~Nathan~Kutz, and S.~L. Brunton, ``Physics-informed dynamic mode decomposition,'' \emph{Proceedings of the Royal Society A}, vol. 479, no. 2271, p. 20220576, 2023.

\bibitem{lee2022structure}
K.~Lee, N.~Trask, and P.~Stinis, ``Structure-preserving sparse identification of nonlinear dynamics for data-driven modeling,'' in \emph{Mathematical and Scientific Machine Learning}.\hskip 1em plus 0.5em minus 0.4em\relax PMLR, 2022, pp. 65--80.

\bibitem{pandey2023robustness}
A.~Pandey and R.~M. Murray, ``Robustness guarantees for structured model reduction of dynamical systems with applications to biomolecular models,'' \emph{International Journal of Robust and Nonlinear Control}, vol.~33, no.~9, pp. 5058--5086, 2023.

\bibitem{matrix_exp_bound_proof_log_norm}
J.~Schmidt, ``G. dahlquist, stability and error bounds in the numerical integration of ordinary differential equations. 85 s. stockholm 1959. k. tekniska h{\"o}gskolans handlingar,'' 1961.

\bibitem{lognorm}
T.~Str{\"o}m, ``On logarithmic norms,'' \emph{SIAM Journal on Numerical Analysis}, vol.~12, no.~5, pp. 741--753, 1975.

\bibitem{aastrom2021feedback}
K.~J. {\AA}str{\"o}m and R.~Murray, \emph{Feedback Systems: An Introduction for Scientists and Engineers}.\hskip 1em plus 0.5em minus 0.4em\relax Princeton University Press, 2021.

\bibitem{antsaklis}
P.~J. Antsaklis and A.~N. Michel, \emph{Linear Systems}.\hskip 1em plus 0.5em minus 0.4em\relax Springer Science \& Business Media, 2006.

\bibitem{pandey2021robustness}
A.~Pandey and R.~M. Murray, ``Robustness guarantees for structured model reduction of dynamical systems,'' in \emph{2021 60th IEEE Conference on Decision and Control (CDC)}.\hskip 1em plus 0.5em minus 0.4em\relax IEEE, 2021, pp. 6920--6927.

\end{thebibliography}
